\documentstyle[seceq,preprint]{ptptex}
\begin{document}

\thispagestyle{empty}
{\baselineskip-4pt
\font\yitp=cmmib10 scaled\magstep2
\font\elevenmib=cmmib10 scaled\magstep1  \skewchar\elevenmib='177
\leftline{\baselineskip10pt\vbox to0pt
   { {\sl\hbox{\hspace{4mm} Osaka  University} }
     {\sl\hbox{{Theoretical Astrophysics}} }\vss}}
\rightline{\baselineskip15pt\rm\vbox to20pt{
\hbox{OU-TAP 64} 
\hbox{YITP-97-41}
\vspace{2mm}
\hbox{\today}\vss}}%
}
\vskip 1cm

\begin{center}
{\large\bf Numerical study on the hydrodynamic instability of 
binary stars in the first post Newtonian approximation of 
general relativity}
\end{center}
%\vspace{5mm}
\begin{center}
Masaru Shibata, Ken-ichi Oohara$^\dagger $, and Takashi Nakamura$^*$ \\
%\vspace{5mm}
{\small\em Department of Earth and Space Science,
Graduate School of Science,
Osaka University,\\ Toyonaka, Osaka 560, Japan}\\
$^\dagger ${\small\em Department of Physics, Niigata University, 
Niigata 950-21, Japan}\\
$^*${\small\em Yukawa Institute for Theoretical Physics, Kyoto University, 
Kyoto 606-01, Japan}
\end{center}
%\baselineskip 7mm
%\vskip 2cm 
\begin{abstract}
We present numerical results on the hydrodynamic stability of 
coalescing binary stars in the first post Newtonian(1PN) 
approximation of general relativity. 
We pay particular attention to the hydrodynamical instability of 
corotating binary stars in equilibrium states 
assuming the stiff polytropic equation of state with 
the adiabatic constant $\Gamma=3$. 
In previous 1PN numerical studies on corotating binary stars 
in equilibrium states, it was found that  along 
the sequence of binary stars as a function of the orbital separation, they 
have the energy and/or angular momentum minima where 
the secular instability sets in, and 
that  with increase of the 1PN correction, the orbital separation 
at these minima decreases while  the angular velocity there 
increases.  In this paper,
to know the location of the innermost stable circular orbit(ISCO),  
 we perform numerical simulations and find
 where the hydrodynamical instability along 
the corotating sequences of binary sets in. 
From the numerical results, we found 
that the dynamical stability limit seems to exist near the 
energy and/or angular momentum minima 
not only in the Newtonian, but also in the 1PN cases. 
This means that the 1PN effect of general relativity increases the 
angular frequency of gravitational waves at the ISCO.
%(version 30 July 1997) 
\end{abstract}
\newcommand{\beq}{\begin{equation}}
\newcommand{\eeq}{\end{equation}}
\newcommand{\beqn}{\begin{eqnarray}}
\newcommand{\eeqn}{\end{eqnarray}}
\newcommand{\pa}{\partial}
\def\bI{\hbox{$\,I\!\!\!\!-$}}
\def\lsim{\mathrel{\mathpalette\Oversim<}}
\def\gsim{\mathrel{\mathpalette\Oversim>}}
\def\Oversim#1#2{\lower0.5ex\vbox{\baselineskip0pt\lineskip0pt%
            \lineskiplimit0pt\ialign{%
          $\mathsurround0pt #1\hfil##\hfil$\crcr#2\crcr\sim\crcr}}}
\def\alt{\lsim}
\def\agt{\gsim}

\break

\section{Introduction}

The laser interferometric gravitational wave detectors 
such as LIGO\cite{LIGO}, VIRGO\cite{VIRGO},
GEO600\cite{GEO} and TAMA300\cite{TAMA} will be in operation 
within four years or so. One of the most important astrophysical sources of
gravitational waves for these detectors is coalescing binary neutron
stars(BNS's) because gravitational waves emitted in the late inspiraling 
phase have frequencies in the sensitive 
region of these detectors, i.e., from 10Hz to 1000Hz.
We will be able to know each
mass and spin of BNS's if we can obtain
an accurate theoretical template for data analysis\cite{CF}.
For that reason, much theoretical effort has been paid to obtain 
a theoretical template as accurate as possible\cite{BDIWW}. 

When the orbital separation of BNS's becomes
a few times of the neutron star(NS) radius, 
the hydrodynamical effect becomes important. In such a post inspiraling phase, 
the wave form of gravitational waves is expected to be sensitive to 
the structure of NS such as the 
relation between the radius and the mass of NS. Thus, if gravitational
waves from such a phase are detected,
we may constrain the equation of state(EOS) of NS\cite{lindblom}
\cite{joan}\cite{LRSLRS}. In particular, it is important to know 
the innermost stable circular orbit(ISCO), where 
the binary will change the orbit from the stable inspiraling orbit to the 
merging one. Since its location will be sensitive to the structure of NS, 
it will bring us an important information on the EOS of NS. 

The location of the ISCO is determined both by (1) the pure 
general relativistic(GR) effect on two spherical bodies 
and by (2) the hydrodynamical effect. 
Recently, Lai, Rasio and Shapiro(LRS) pointed out the 
importance of the latter effect\cite{LRSLRS}: 
They showed that when the orbital 
separation of BNS's becomes sufficiently small, but larger than 
the approximate radius of the ISCO for two spherical bodies $\sim 6GM/c^2$ 
where $M$ is the total mass of the system, 
each star of binary is significantly 
deformed by the tidal force from the companion star. Then, an 
additional tidal field due to the deformation of each star is generated and 
as a result, the circular orbit of BNS's becomes hydrodynamically unstable 
before two stars come into contact even in 
the Newtonian case if the EOS of NS is stiff enough. 

Such a tidal effect will be sensitive to the structure of NS because 
the degree of the tidal deformation depends on it. Since 
general relativity plays an important role in determining the 
structure of NS, we can expect that this is also the case 
in determining the location where 
the hydrodynamical stability due to the tidal force sets in. 
To know the effect of general relativity, as a first step, 
Shibata\cite{shibapn}\cite{shibapnb} numerically obtained 
the equilibrium state of corotating binary of equal mass 
in the 1PN approximation and investigated the stability of such binaries. 
The purpose of his works was to clarify the 1PN effect of general 
relativity in BNS's. The main conclusions he obtained are 
(a) along the equilibrium sequence, there exist the energy and/or 
angular momentum minima where the secular instability of binary 
sets in\cite{LRSLRS}, 
and (b) the orbital separation at the 
minima decreases due to 1PN correction, and 
as a result, the orbital angular velocity there increases. 

As LRS showed for the Newtonian case, 
the secular instability limit does not coincide with the dynamical one 
for corotating binary, and the orbital separation of the latter one 
will locate slightly inside the former one\cite{LRSLRS}. 
The ISCO is defined as the dynamical instability limit, 
which is found only by performing numerical simulation. Hence, 
in order to determine the
dynamical instability limit along the sequence of corotating binary, 
 we should perform numerical simulations and judge the dynamical stability. 
In this paper we will show that  the hydrodynamical instability 
occurs  near the energy and/or angular momentum minima not only 
in the Newtonian cases but also in the 1PN cases. 

This paper is organized as follows. In \S 2, we show the 
1PN hydrodynamic equations to be solved which are slightly different 
from previous ones\cite{bds}: In particular, we change the form of 
the energy equation. 
In \S 3, we briefly describe the method for numerical simulation. 
In \S 4, numerical results 
are shown: In this paper, we pay attention to binaries of the polytropic EOS 
of the adiabatic index $\Gamma=3$ for which the energy and angular 
momentum minima for equilibrium sequences were clearly obtained 
for both the Newtonian and 1PN cases in previous papers\cite{shibapn}
\cite{shibapnb}. 
We show numerical results and discuss  the dynamical stability of binaries around those minima. 
\S 5 is devoted to summary.  
Throughout this paper, $G$, $c$, and $M_{\odot}$ denotes 
the gravitational constant, the speed of light and the solar mass. 

\section{Basic equations}

In the 1PN approximation with the standard 
gauge condition, the line element is written as\cite{asada}
$$
ds^2=g_{\mu\nu}dx^{\mu}dx^{\nu}=
-\alpha^2c^2dt^2+{2 \over c^2}\beta_i dx^idt+\psi^4 \tilde \gamma_{ij}
dx^idx^j
$$
where
\beqn
\alpha&=&1-{U_* \over c^2}+{1 \over c^4}\left({U_*^2 \over 2}+X_*\right)
+O(c^{-6}),\\
\psi&=&1+{U_* \over 2c^2}+O(c^{-4}),\\
\tilde \gamma_{ij}&=&\delta_{ij}+O(c^{-4}).
\eeqn
Here, $\beta_i$ is calculated from 
\beq
\beta_{i}=-{7 \over 2}P_{*i}
+{1 \over 2}\Bigl( x^j P_{*j,i}+\chi_{,i}\Bigr).
\eeq
Equations for gravitational potentials, 
$U_*$, $X_*$, $P_{*i}$ and $\chi$, are shown below. 

The energy-momentum tensor is 
\beq
T^{\mu\nu}=\rho\left(c^2+\varepsilon+{P \over \rho}\right) u^{\mu}u^{\nu}
+Pg^{\mu\nu},
\eeq
where $\rho$, $\varepsilon$, $P$ and $u^{\mu}$ are the baryon density, 
the internal energy, the pressure, and the four velocity, respectively. 
In the 1PN approximation, components of four velocity are written as 
\beqn
u^0&=&1+{1 \over c^2}\Bigl\{{1 \over 2}v^2+U_* \Bigr\}+O(c^{-4}),\\
u_0&=&-\biggl[ 1+{1 \over c^2}\Bigl\{{1 \over 2}v^2-U_*\Bigr\}+O(c^{-4})
\biggr],\\
u^i&=&{v^i \over c}
\biggl[1+{1 \over c^2}\Bigl\{{1 \over 2}v^2+U_*\Bigr\}+O(c^{-4})\biggr],\\
u_i&=&{v^i \over c}+{1 \over c^3}\Bigl\{\beta_i+v^i \Bigl({v^2 \over 2}+3U_* 
\Bigr)\Bigr\}+O(c^{-5}),
\eeqn
where 
\beq
v^i = {u^i \over u^0}~~{\rm and} ~~v^2=v^iv^i.
\eeq
As the EOS, we assume the polytropic one as 
\beq
P=(\Gamma-1) \rho \varepsilon. 
\eeq

The  continuity equation is 
\beq
\nabla_{\mu} (\rho u^{\mu})=0~.
\eeq
 Using the variable $\rho_*=\rho \alpha u^0 \psi^6$, we rewrite the
continuity equation as
\beq
{\partial \rho_* \over \partial t}
+{\partial (\rho_* v^i) \over \partial x^i}=0~. 
\eeq
In the 1PN approximation, $\rho_*$ is 
\beq
\rho_*=\rho\biggl\{1+{1 \over c^2}\Bigl({v^2 \over 2}+3U_* \Bigr)
+O(c^{-4}) \biggr\} .
\eeq
From the conservation of the energy momentum tensor, 
\beqn
\nabla_{\mu} T^{\mu}_{~\nu}
&&={1 \over \alpha \psi^6}{\partial \over \partial x^{\mu}}
\biggl\{\alpha \psi^6 \Bigl(\rho c^2+\rho \varepsilon+P\Bigr) u^{\mu}u_{\nu}
\biggr\} \nonumber \\
&& \hskip 1cm -{1 \over 2}\Bigl(\rho c^2+\rho \varepsilon+P\Bigr)
u^{\mu}u^{\sigma}g_{\mu\sigma,\nu}+P_{,\nu} \nonumber \\
&&=0, 
\eeqn
we obtain eqautions for the Euler and the entropy conservation  as 
\beqn
&&{\partial (\rho_* \hat u_i) \over \partial t}
+{\partial (\rho_* \hat u_i v^j) \over \partial x^j} 
=-\alpha \psi^6 P_{,i}-\rho_* \alpha \hat u^0 \alpha_{,i}
+\rho_* \hat u_j \beta^{j}_{~,i}
-{\rho_* \over 2\hat u^0}\hat u_j \hat u_k \gamma^{jk}_{~~,i}\hskip 1cm\\
%%%%%%%%%
&&{\partial e_* \over \partial t}
+{\partial (e_* v^j) \over \partial x^j}=0,
\eeqn
where 
\beqn
\hat u_i && \equiv \biggl\{ 1 +{1 \over c^2}
\Bigl( \varepsilon+{P \over \rho} \Bigr)\biggr\} u_i , \\
%%%%%%%%%%%%%%%
\hat u^0 && \equiv \biggl\{ 1 +{1 \over c^2}
\Bigl( \varepsilon+{P \over \rho} \Bigr)\biggr\}u^0 , \\
%%%%%%%%%%%%%%%
e_* && \equiv (\rho \varepsilon)^{1/\Gamma}\alpha u^0 \psi^6.  
\eeqn

%We note that the equation for $e_*$ may be regarded as the 
%GR entropy conservation equation. 
In the 1PN approximation, the Euler equation becomes
\beqn
&&{\partial (\rho_* \hat u_i) \over \partial t}
+{\partial (\rho_* \hat u_i v^j) \over \partial x^j} 
= -\Bigl( 1+{2U_* \over c^2} \Bigr)P_{,i}
+\rho_*\biggl[ \Bigl\{ 1 
+{1 \over c^2}\Bigl(\varepsilon+{P \over \rho}
+{3v^2 \over 2}-U_* \Bigr)\Bigr\}U_{*,i}\nonumber \\
&&\hskip 3cm
-{X_{*,i} \over c^2}+{v^j\beta_{j,i} \over c^2}\biggr]+O(c^{-4})~,
\eeqn
where  
\beq
\hat u_i = v^i\biggl\{1+{1 \over c^2}
\Bigl(\varepsilon+{P \over \rho}+{v^2 \over 2}+3U_* \Bigr)\biggr\}
+{\beta_i \over c^2}+O(c^{-4})~.
\eeq
The pressure is calculated from 
\beq
P=(\Gamma-1)e_*^{\Gamma}\biggl\{ 
1+{1 \over c^2}\Bigl({v^2 \over 2}+3U_* \Bigr)\biggr\}^{-\Gamma}. 
\eeq 

We notice that the form of the 1PN hydrodynamic equations, in particular 
the entropy equation, shown above 
are different from those of Blanchet, Damour, and 
Sch\"afer\cite{bds} although in both formalisms, 
every 1PN term is taken into account consistently.
Thus, equations adopted in this paper are different 
from those used in a previous paper\cite{on}. 

Equations for various potentials 
$U_*$, $X_*$, $P_{*i}$ and $\chi$ are derived from the Einstein equation as 
\beqn
&&\Delta U_*=-4\pi G\rho_* ,\\
&&\Delta P_{*i}=-4\pi G\rho_* v^i,\\
&&\Delta X_*=4\pi G\rho_* \biggl({3v^2 \over 2}-U_*+
\varepsilon+{3P \over \rho}\biggr),\\
&&\Delta \chi =4\pi G\rho_* v^i x^i .
\eeqn
Note that the definitions of the gravitational potentials are based on 
$\rho_*$, not on $\rho$. Thus, they are different quantities from 
those defined in obtaining the equilibrium 
state\cite{shibapn}\cite{shibapnb} in which we need to use 
the gravitational potentials defined with respect to 
$\rho$. (For example, the Newtonian potential $U$ which satisfies 
$\Delta U=-4\pi\rho$ differs from $U_*$. ) 

We use the following forms of the continuity, Euler, and the entropy  
equations as the basic equations in the Newtonian case:
\beqn
&&{\pa \rho \over \pa t}+{\pa (\rho v^j) \over \pa x^j}=0,\\
&&{\pa (\rho v^i )\over \pa t}+{\pa (\rho v^i v^j) \over \pa x^j}=
-{\pa P \over \pa x^i}+\rho {\pa U \over \pa x^i},\\
&&{\pa e \over \pa t}+{\pa (e v^j) \over \pa x^j}=0, 
\eeqn
where $e=(\rho \varepsilon)^{1/\Gamma}=[P/(\Gamma-1)]^{1/\Gamma}$. 
Thus, we change the form of the energy equation from that adopted 
in previous papers\cite{non}\cite{sno}. 
We emphasize that this choice 
improves accuracy on the conservation of the energy and angular momentum 
in numerical simulation compared with our previous works. 

Before closing this section, we define the following quantities 
correct up to 1PN order:
\begin{itemize}
\item Conserved mass and Newtonian mass
\beq
M_*=\int \rho_* dV~~~~~{\rm and}~~~~~M=\int \rho dV.
\eeq
Note that for the Newtonian case, $M_*=M$. 
\item The energy: In the definition based on $\rho$, 
\beqn
E=\int \rho \biggl\{\varepsilon+{v^2 \over 2}-{1 \over 2}U
+{1 \over c^2}\biggl(&&{5 \over 8}v^4+{5 \over 2}v^2U+{1 \over 2}\beta_i v^i
+\varepsilon v^2+{P \over \rho}v^2 \nonumber \\
&&+2\varepsilon U-{5 \over 2}U^2 \biggr)
\biggr\}dV,
\eeqn
while in the definition based on $\rho_*$, 
\beqn
E&&=\int \rho_* \biggl\{\varepsilon+{v^2 \over 2}-{1 \over 2}U_*
+{1 \over c^2}\biggl(
{3 \over 8}v^4+{3 \over 2}v^2U_*+{1 \over 2}\beta_i v^i
+{1 \over 2}\varepsilon v^2+{P \over \rho}v^2 \nonumber \\
&& \hskip 7cm -\varepsilon U_*+{1 \over 2}U_*^2 \biggr)
\biggr\}dV \nonumber \\
&& \equiv E_*. 
\eeqn
\item The angular momentum around the rotation axis : In the 
definition based on $\rho$, 
\beq
J=\int \rho \biggl [ (-yv^x+x v^y)\biggl\{1+{1 \over c^2}
\biggl(v^2+6U+\varepsilon+{P \over \rho}\biggr) \biggr\}
+{1 \over c^2}(-y\beta_x+x \beta_y)\biggr]dV,
\eeq
while in the definition based on $\rho_*$, 
\beq
J=\int \rho_* (-y\hat u_x+x \hat u_y) dV\equiv J_*. 
\eeq
Note that we adopt the $z$-axis as the rotation axis in this paper. 
Here, we also notice 
that numerical values $E$ and $E_*$ or $J$ and $J_*$ are, respectively,  
slightly and systematically 
different because some terms of $O(c^{-4})$ 
are implicitly  included in $E_*$ and $J_*$. 
However, the difference is small and not important in the following
discussions. In the following, we use $J_*$ as the angular momentum in numerical simulations, while 
we use the definition based on $\rho$ as the angular momentum of the equilibrium sequence. 
\item Center of mass of each star
\beq
x_g^i={1 \over M_*} \int_{\rm each~ star} \rho_* x^i dV.
\eeq
From $x_g^i$, we define the coordinate separation of the orbital radius 
as $r_g=2(x_g^i x_g^i)^{1/2}$. 
\item The quadrupole moment of system
\beq
I_{ij}=\int \rho_* x^i x^j dV. 
\eeq
We also define 
\beq
I_{RR}\equiv I_{xx}+I_{yy},~~~~~{\rm and}~~~~~\bI_{ij}=I_{ij}-{1 \over 3}
\delta_{ij}I_{kk},
\eeq
where $\delta_{ij}$ is the Kronecker's delta. 
\end{itemize}

\section{Numerical method}

In numerical computation, we solve (1) the 1PN hydrodynamic 
equations and (2) the Poisson equations for various potentials. 
In the following, we mention the numerical methods we adopt briefly. 

Our treatment for the advection term in the hydrodynamic equations 
is the same as that adopted in previous papers\cite{non}\cite{on}\cite{sno}. 
Although the method is the same, 
the energy equation we adopt is different from the previous one. 
As a result, accuracy on 
the conservation of the angular momentum and the energy of 
the system is significantly improved.\footnote{
We here note that using the entropy equation adopted above, the 
entropy is conserved. If the shock is formed, however, 
the entropy should not be conserved. 
Hence, we must introduce the artificial viscosity term in the 
Euler and entropy equations to express the shock accurately. 
In this paper, however, the shock formation does not play an important 
role, so that we do not add such terms. }
As for the time evolution, we use the second order Runge-Kutta method. 
Thus, the numerical code for solving the hydrodynamic equations is 
second order accurate both in time and space. (Actually, we checked, for 
example, that accuracy of the 
conservation of the total energy is second order convergent 
changing the grid number.) 

Since we use the stiff EOS, density near the surface of each star decreases 
steeply. With a finite resolution of numerical grid, 
one cannot accurately represent the density decline 
near the surface. As a result, the pressure gradient 
often happens to become too 
steep, and the velocity near the surface becomes unphysically too large. 
To avoid that, when the density $\rho_*$ is less than 
$\rho_{\rm crit} \simeq 
2\times 10^{-4}\rho_{\rm max}$, where $\rho_{\rm max}$ is
the maximum of $\rho_*$, 
we artificially suppress the linear momentum $\hat u_i$(or $v^i$ 
for the Newtonian case) near the surface 
by a factor $\rho_* / (\rho_* + f\rho_{\rm max})$, where 
$f$ is a small factor and we set as $\simeq 4 \times 10^{-5}$. 
This technique has been used in our series of papers\cite{non}\cite{on}
\cite{sno}.

To solve 6 Poisson equations for $U_*$, $X_*$, $P_{*x}$, 
$P_{*y}$, $P_{*z}$, and $\chi$, we use the ICCG method as we have done 
in previous papers\cite{non}\cite{on}\cite{sno} 
imposing the boundary conditions at outer grids as 
\beqn
&&U_* \rightarrow {GM_* \over r}+O(r^{-3}),\\
&&X_* \rightarrow -{G \over r}\int \rho_* 
\biggl({3v^2 \over 2}-U_*+\varepsilon+{P \over \rho}\biggr) dV+O(r^{-3}),\\
&&P_{*x} \rightarrow {Gx \over r^3}\int \rho_* x v^xdV
+{Gy \over r^3}\int \rho_* yv^x dV+O(r^{-4}),\\
&&P_{*y} \rightarrow {Gx \over r^3}\int \rho_* x v^ydV
+{Gy \over r^3}\int \rho_* yv^y dV+O(r^{-4}),\\
&&P_{*z} \rightarrow {Gz \over r^3}\int \rho_* z v^zdV+O(r^{-4}),\\
&&\chi \rightarrow -{G \over r}\int \rho_* v^i x^idV+O(r^{-3}).
\eeqn
Thus, we use three types of numerical implementations for solving Poisson 
equations for ($U_*$, $X_*$, $\chi$), 
($P_{*x}$, $P_{*y}$), and $P_{*z}$, respectively. 
Accuracy of these Poisson solvers has been checked to be less 
than $0.1\%$ and a result for the case of a small grid number 
is shown in \cite{ah}. 

Throughout this paper, we assume the symmetry with respect to 
the equatorial plane. Thus, we also impose boundary conditions 
at the equatorial plane as 
\beqn
&&\rho_*(-z)=\rho_*(z),~~~e_*(-z)=e_*(z),~~~v^A(-z)=v^A(z),\nonumber \\
&&v^z(-z)=-v^z(z),~~~U_*(-z)=U_*(z),~~~X_*(-z)=X_*(z),\nonumber \\
&&\chi(-z)=\chi(z),~~~P_{*A}(-z)=P_{*A}(z),~~~P_{*z}(-z)=-P_{*z}(z),
\eeqn
where $A=x$ or $y$. 
Grid number we adopt is $(N_x, N_y, N_z)=(141,141,71)$, and grid spacing 
is chosen in order that 
the grid covers the minor axis of each star at initial state 
by at least 15 grid points. 
Numerical computations were performed on FACOM VX4 in the data processing 
center of National Astronomical Observatory in Japan. 

FACOM VX4 has four parallel processors, so that four different 
procedures can be done on separate processors at the same time. 
To solve the Poisson equations in 
the 1PN case, we make use of this property: In each time step, 
we solve the Poisson equations for $U_*$ and $X_*$ on processor 1, 
for $P_{*x}$ and $P_{*y}$ on processor 2, for $P_{*z}$ on processor 3, 
and for $\chi$ on processor 4, while the hydrodynamic equations are solved 
only on processor 1. Since computation of 
solving the Poisson equations is the most time-consuming part in the 
numerical implementations, 
the CPU time required reduces by a factor of $\sim 3$ in the 1PN case 
using this simple technique. On the other hand, we 
only use one processor in the Newtonian case because only one 
Poisson equation is required to be solved. 
In a typical computation for one model, $20000\sim 30000$ time 
steps were necessary, 
and computation of 20000 time steps took 
about 30 CPU hours for the Newtonian case, 
and about 65 CPU hours for the 1PN case.

\section{Numerical results}

We assume $\Gamma=3$ as the adiabatic index 
because we are interested in 
BNS's   which are composed of a stiff EOS, i.e., 
$\Gamma = 2\sim 3$\cite{st}. 
As initial conditions of binary, we adopt equilibrium states of binary 
corotating around the $z$-axis. When we obtain such a solution, we set the 
EOS as 
\beq
P=K \rho^{\Gamma},
\eeq
where $K$ is the polytropic constant. 
Following a previous paper\cite{shibapn}, for convenience, we set $K$ as 
\beq
K=2.524 G r_0^5 M_{\odot}^{-1}.
\eeq
Then, the radius of each star of binary in the Newtonian limit is 
$r_0(M/M_{\odot})^{0.2}\equiv a_0$. 
In the following, $M_*$ denotes the conserved mass of each star 
of binary and is set as $1.4M_{\odot}$. 
In this paper, we consider two cases, 
the Newtonian case(i.e., $c \rightarrow \infty$ in the PN approximation) 
and the 1PN case of $r_0=40R_s$ where $R_s=GM_{\odot}/c^2$. 
We note that for $r_0=40R_s$, the compactness of each star is  
$C_{\rm s}\equiv GM_*/a_* c^2 \simeq 0.033$, where 
$a_* \equiv r_0(M_*/M_{\odot})^{0.2}$(note that $a_*=a_0$ 
in the Newtonian case), and 
this value is too small to represent realistic BNS's 
which we are most interested in. 
However, the PN approximation is valid only for weak gravity, i.e., for 
small compactness, and it is meaningless to apply it for large compactness. 
We emphasize that 
the main aim of this paper is to solidly 
know the 1PN effect which will play an important role in BNS's. 

In figs.1, we show the energy and the angular momentum 
of corotating binary in equilibrium states as a function of
the orbital separation $r_g/a_*$ 
for the 1PN case of $r_0=40R_s$(filled circles) and the 
Newtonian case(open circles). 
As shown in previous papers\cite{shibapn}\cite{shibapnb}, 
there exist the minima along the sequences 
of the energy and angular momentum, and the separation at the 
minima becomes small with increase of the 1PN effect: 
For the Newtonian case, 
the separation is $r_g \simeq 3.08a_0$, and for the 1PN case of 
$r_0=40R_s$, it is $r_g \simeq 2.94a_*$. 

We will perform numerical simulations using 
some of such equilibrium states 
near the energy and/or angular momentum minima as initial conditions. 
Among equilibrium states along each sequence, 
we choose four states as initial conditions: 
they are  N5, N6, N7, and N8 along 
the Newtonian sequence, and  PN5, PN6, PN7, and PN8 along the 1PN sequence, 
respectively. Models N6 and PN6 are equilibrium states at the 
nearest location from the energy and/or angular momentum minima. 

In figs.2(a) and (b), we show the evolution of $I_{RR}$ for the Newtonian and 
1PN cases, respectively. Hereafter, we use the unit of time 
as $4\pi \sqrt{a_*^3/GM_*}$, and 
in this unit, the orbital periods of equilibrium states 
are $\simeq 1.82$ for N5, $1.88$ for N6, 
$1.95$ for N7,  $2.03$ for N8, 
$1.74$ for PN5, $1.80$ for PN6,
$1.87$ for PN7, $1.95$ for PN8, respectively. 
On the other hand, the vertical axis is shown in units of 
$M_*a_*^2$(i.e., $I=I_{RR}/(M_* a_*^2)$). 
This means that when $I \alt 2$, the binary begins to merge. Hereafter, 
solid, dotted, dashed and dotted-dashed lines are used to show 
quantities for 
models N5(PN5), N6(PN6), N7(PN7), and N8(PN8), respectively. 

Figs.2 show that equilibrium models 
N5, N6, PN5 and PN6 seem hydrodynamically unstable; two stars 
merge within about one orbital period. On the other hand, models 
N7, N8, PN7, and PN8 seem hydrodynamically stable because in these 
binaries, two stars do not come into merging in the 
dynamical time scale(i.e., the orbital period). 
We note that irrespective of the hydrodynamical stability, 
the orbit of binaries shrinks due to 
dissipation of the angular momentum. This occurs because of a finite 
resolution(i.e., finite accuracy) both in time and space 
in the numerical simulation, and 
is the reason why $I$ decreases 
gradually for stable binaries such as N7, N8, PN7, and PN8. 

In fig.3, we show the 
angular momentum as a function of time for the Newtonian 
and 1PN cases( note that for the 1PN case, the angular momentum is $J_*$).  
Here, the unit of the angular momentum is $G^{1/2}M_*^{3/2}a_*^{1/2}$, and 
the unit of the time is the same as that in figs.2. 
Notice that a small fraction of matter carrying 
the large angular momentum is ejected outside the numerical grids 
in  model PN5 for $t \agt 2.6$, and in model N5 for $t \agt 2.9$ so
that the angular momentum is decreasing. 
Fig.3 shows that accuracy on 
the conservation of the angular momentum is fairly good in all 
simulations, but it is not perfect: In one orbital period, the 
angular momentum is lost by about $\sim 0.1 \%$. 

Even for the angular momentum loss, equilibrium models 
N7, N8, PN7, and PN8 do not proceed to merging in the dynamical 
time scale. Thus, these binaries seem obviously stable. 
On the other hand, we cannot completely deny the possibility 
that a small loss of the angular momentum by such numerical effects 
leads models N5, N6, PN5 and PN6 to unstable orbits. However, the loss of the 
angular momentum is not so large up to the contact of these binaries. 
According to an approximate analysis by 
LRS(see, for example, fig.1 in \cite{LRSLRS}), 
the hydrodynamically stable corotating binary 
will move to another stable binary which is not corotating state 
even when it slightly loses the 
angular momentum(unless its loss is so large). 
Hence, we judge that the hydrodynamical instability will really occur 
when the orbital separation of binary becomes as small as that of 
N5, N6, PN5 and PN6 or slightly smaller than these, 
i.e. very near or slightly inside the energy and/or 
angular momentum minima along the sequences of corotating binaries, 
$r_g \sim 3.08a_0$ for the Newtonian 
case, and $r_g \sim 2.94a_*$ for the 1PN case of $C_{\rm s} \simeq 0.033$. 

We note the following fact: 
As for the Newtonian study on the determination of 
the dynamically stability limit 
for the $\Gamma=3$ case, there have been two independent works by
Rasio and Shapiro(RS)\cite{rs} and New and Tohline(NT)\cite{nt}, and 
their results slightly disagree: 
According to RS, the dynamical instability along 
the corotating sequence sets in at $r_g \simeq 2.97a_0$, while according to NT, 
it sets in at $r_g \simeq 3.08a_0$. 
The hydrodynamic code which RS used in their numerical 
simulations is a SPH one, while that 
used by NT is the Eulerian one. In our present simulation, we use 
the Eulerian code, and this may be the reason why 
the present results support the work by NT. 
Currently, however, the reason for this slight difference is not clear. 

In fig.4, we show the maximum density of the system 
as a function of time for the Newtonian and 1PN cases, respectively. 
The density is shown in units of $M_{*}(4\pi a_*^3/3)^{-1}$. 
Fig.4 shows that apart from a small oscillation,
the maximum density is almost constant for the case where
two stars of the binary do not merge.
We notice that the maximum density for the 1PN configuration are larger
than that for the Newtonian one for a given angular frequency. 
This is a manifestation 
of the GR effect by which each star of binary 
is forced to be more compact than that in the Newtonian case. 
It should be also noted that along not only the Newtonian equilibrium 
sequence, but also the 1PN one, 
the central density decreases with decrease of 
the orbital separation\cite{tani}\cite{kip}. 

In figs.5 and 6, we show the 
evolution of the density profile in the equatorial plane 
for models N7 and PN7. 
Note that $x$- and $y$-axes are shown in units of $a_*$, 
outermost grid is placed at 4.09 in this unit, and 
the orbital rotation is counterclockwise. 
Figs.5 and 6 show that for more than one orbital period, 
the density profile does not change so much apart from a small deformation 
in the outer region of each star. 
Thus, in our present numerical simulation, the 
density profile of equilibrium state 
can be kept fairly well for hydrodynamically stable models.

In figs.7 and 8, 
we show the evolution of the density profile in the equatorial plane 
for models N5 and PN5, i.e., dynamically unstable models, respectively. 
In both Newtonian and 1PN cases, the evolution is similar to that 
obtained by RS\cite{rs}:
In about one orbital period, two stars merge due to the hydrodynamic 
instability. 
After they merge, spiral arms are formed and they transport 
the angular momentum outward. 
In the final stage, the spiral arms widen and merge together. 
In the central region, an ellipsoidal core is formed. Since 
radiation reaction of gravitational waves is not taken into 
account in this paper, the ellipsoidal figure will be kept forever. 

Finally, we show the wave form and the luminosity  
of gravitational waves for merging phase of binary. 
Although we are able to calculate the 1PN wave form 
and luminosity using the Blanchet, Damour, 
and Sch\"afer formalism\cite{bds}, 
in this paper we simply show the Newtonian ones.  
In figs.9(a) and (b), we show 
$ + $ and $\times $ modes of gravitational waves 
observed along the $z$-axis for models PN5(solid lines) and N5(dotted lines). 
$ + $ and $\times $ modes are defined as 
\beq
r_o h_+={G \over c^4}
{d^2 \over dt^2}\biggl(I_{xx}-I_{yy}\biggr),~~~~~{\rm and}~~~~~
r_o h_{\times}={2G \over c^4}{d^2 \over dt^2}I_{xy},
\eeq
where $r_o$ denotes the distance from the source to an observer on
the $z$-axis. 
In fig.9(c), we show the luminosity defined as 
\beq
{dE \over dt}={G \over 5c^5}{d^3 \bI_{ij}\over dt^3}
{d^3 \bI_{ij}\over dt^3}. 
\eeq

Since the orbital period of model PN5 is shorter than that of N5, 
the wave length of gravitational waves before merging 
for PN5 is shorter than that of N5. Also, the 
each star of model PN5 is more compact than that of N5, so that the 
maximum amplitude of the luminosity for PN5 is larger than that 
for N5. Although these differences exist between PN and Newtonian results, 
overall shapes of the wave form and luminosity in the PN approximation 
are essentially the same as those in the Newtonian case(see 
also figs.10 and 11 in ref.\cite{rs}): 
As the orbit shrinks due to the hydrodynamical 
instability, the amplitude and the luminosity of gravitational waves 
gradually rise. When two stars of the binary come into contact, they 
become the maximum, and after that, they gradually fade away. 
Since the merged object becomes an ellipsoid in the final state and 
radiation reaction of gravitational waves is not included in this 
simulation, the wave form will settle down to a stationary one.

Before closing the section, we notice that 
evolution of the merged object and gravitational waves from it 
considerably depend on the velocity field of pre-merging state of binary: 
In the case where each star of 
binary does not have spin such as the irrotational Roche-Riemann binary 
which is regarded as one of the most realistic situations for BNS\cite{irre}, 
we have learned from numerical simulations 
that the large spiral arm is not formed\cite{sno}\cite{max}. 
Also, when radiation reaction of gravitational waves is included and 
as a result, pre-merging binary has an approaching velocity, the 
merged object radially oscillates and second and third peaks of 
the gravitational wave luminosity, which in the present result do 
not appear, can be seen\cite{sno}\cite{joan}\cite{max}. 
Thus, we should keep in mind that the present results on the evolution of 
merged object and gravitational waves from it are inherent for 
evolution of dynamically unstable corotating binary 
without radiation reaction.

\section{Summary}

In this paper, we have shown numerical results of 
 the hydrodynamical stability of 
corotating binary in equilibrium states in the 
1PN approximation of general relativity. 
We found that the hydrodynamical instability sets in near the 
energy and/or angular momentum minima along the equilibrium sequence. 
This means that we may approximately 
determine the location of the ISCO from the energy and/or 
angular momentum minima along sequences for corotating binary. 

As shown in previous papers, the angular velocity at those minima 
changes with the first GR correction approximately 
as\cite{shibapn}\cite{shibapnb} 
\beq
\Omega=\Omega_{\rm N}\biggl(1+C_{\rm PN}{GM_* \over a_* c^2}\biggr),
\eeq
where $\Omega_{\rm N}$ is the angular velocity 
at the minima for the Newtonian case 
and $C_{\rm PN}$ is a constant $\sim 1.1$ for $\Gamma=3$ and 
$\sim 2.5$ for $\Gamma=2$. For the polytropic EOS 
of $\Gamma=3$\cite{shibapn} and $2$\cite{shibapnb}, 
\beqn
\Omega_{\rm N}
&&\simeq (0.267 \pm 0.002)\sqrt{{GM \over a_0^3}}~~~{\rm for}~ \Gamma=3,\\
&&\simeq (0.297 \pm 0.002)\sqrt{{GM \over a_0^3}}~~~{\rm for}~ \Gamma=2,
\eeqn
and the frequency of gravitational waves is calculated as 
\beqn
f_{\rm N} \equiv {\Omega_{\rm N} \over \pi} 
&&\simeq630\biggl({M \over 1.4M_{\odot}}\biggr)^{1/2}
\biggl({15 {\rm km} \over a_0}\biggr)^{3/2}{\rm Hz}~~~{\rm for}~ \Gamma=3,\\
&&\simeq700\biggl({M \over 1.4M_{\odot}}\biggr)^{1/2}
\biggl({15 {\rm km} \over a_0}\biggr)^{3/2}{\rm Hz}~~~{\rm for}~ \Gamma=2,
\eeqn
where $M$ and $a_0$ denote the Newtonian mass and 
radius of each star of binary, respectively. 
Since the EOS of NS has the adiabatic index of $2 \sim 3$, 
$f_{\rm N}$ may be approximated by 
\beq
f_{\rm N} \simeq 600-700 \biggl({M \over 1.4M_{\odot}}\biggr)^{1/2}
\biggl({15 {\rm km} \over a_0}\biggr)^{3/2}{\rm Hz}.
\eeq
Thus, in the case of corotating binary, 
the frequency of gravitational waves 
at the ISCO for a realistic BNS of radius 10$\sim$15km(i.e., 
$C_{\rm s} =0.14 \sim 0.2$) will become about $150 \sim 300$Hz higher 
due to the 1PN correction of general relativity. 

We performed numerical simulations 
only including the 1PN correction, i.e., the lowest order GR correction, 
in this paper, but it is not sufficient 
in order to obtain the ISCO for realistic BNS's accurately 
because BNS's are highly GR objects($C_{\rm s} \sim 0.2$). 
We need to take into account the fully GR term to 
obtain an accurate result. We, however, emphasize that 
the conclusion  in this paper will hold for binaries of small 
compactness(say $C_{\rm s} \alt 0.05$). This means that when we 
perform fully GR simulation, the numerical code must reproduce 
the present result. Thus, the present paper will be a useful 
guideline in checking fully GR calculations. 

\vskip 5mm
\begin{center}
{\bf Acknowledgments }
\end{center}
\vskip 5mm

Numerical computations were mainly performed on FACOM VX4 in
data processing center of National Astronomical Observatory in Japan.
This work was
in part supported by a Grant-in-Aid of Ministry of Education, Culture,
Science and Sports (08NP0801, 08237210 and 09740336).

\vspace{1cm}

\begin{center}
{\large\bf Figure captions}
\end{center}
\vskip 5mm

\begin{description} 
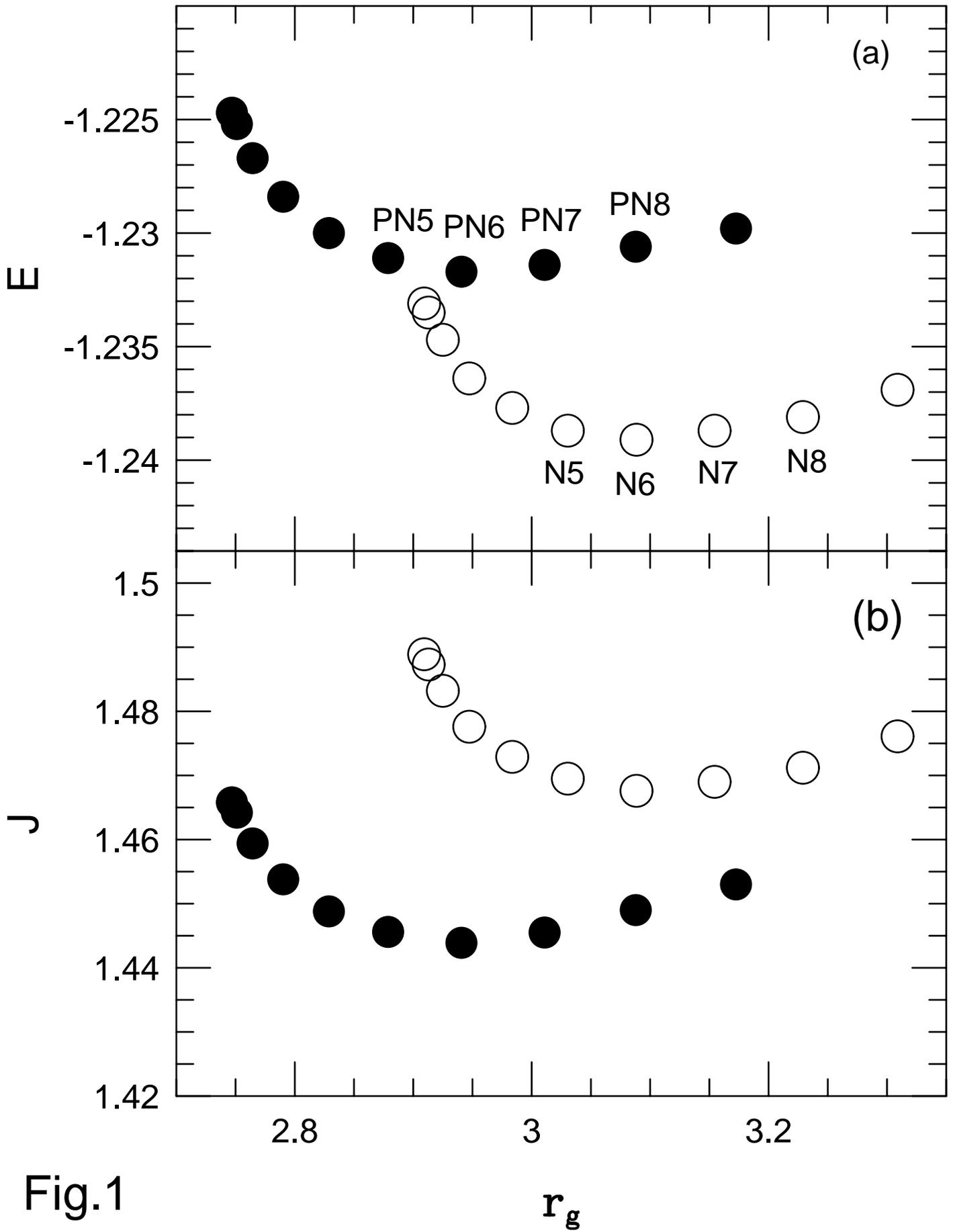
\item{\rm Fig.1} The energy(a) and the angular momentum(b) 
of corotating binary in equilibrium states as a function of
the orbital separation $r_g/a_*$ 
for the 1PN case of $r_0=40R_s$(filled circles) and 
the Newtonian case(open circles).
The energy and the angular momentum are shown in units of 
$GM_*^2/a_*$ and $G^{1/2}M_*^{3/2}a_*^{1/2}$, respectively. 
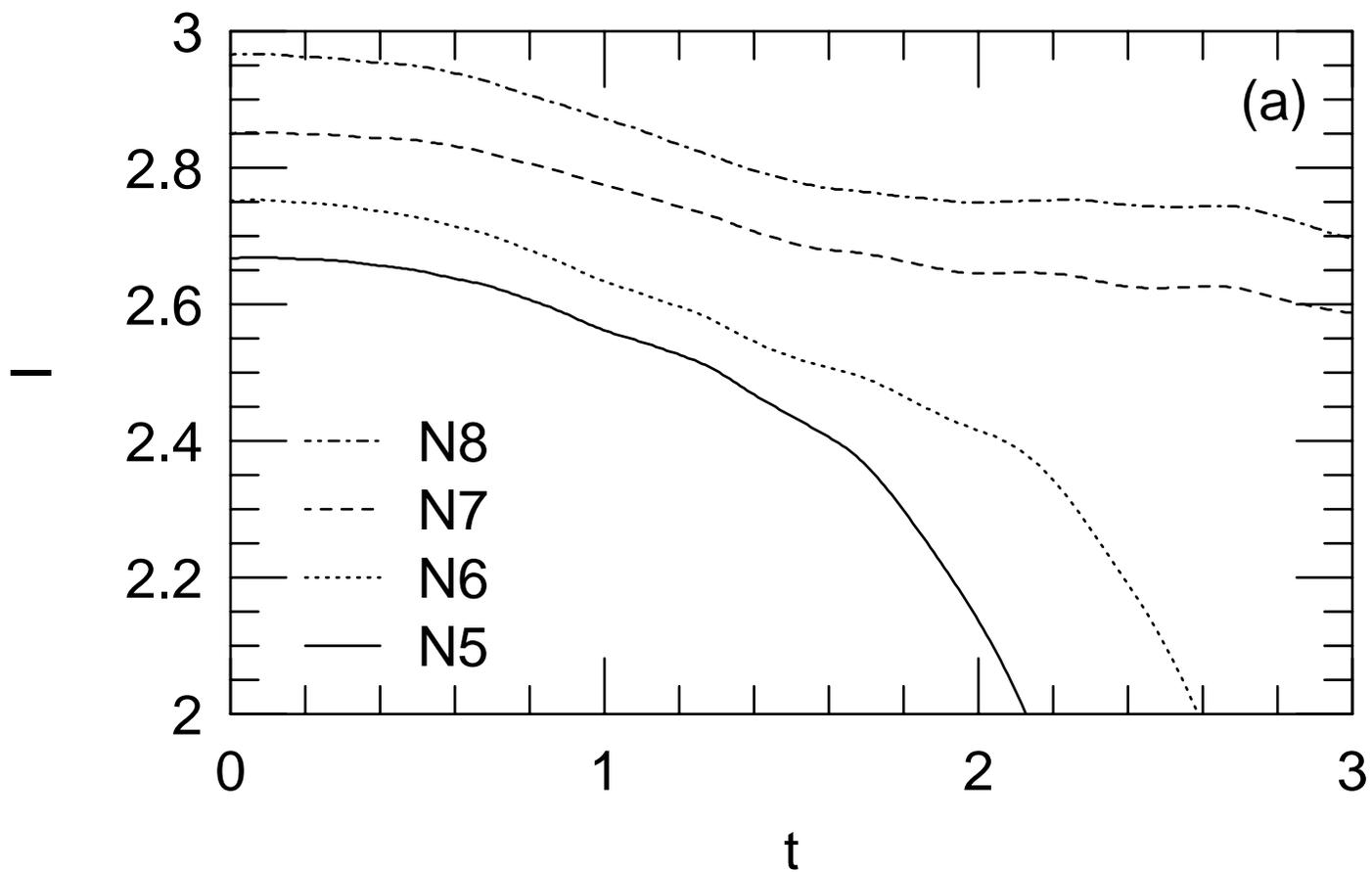
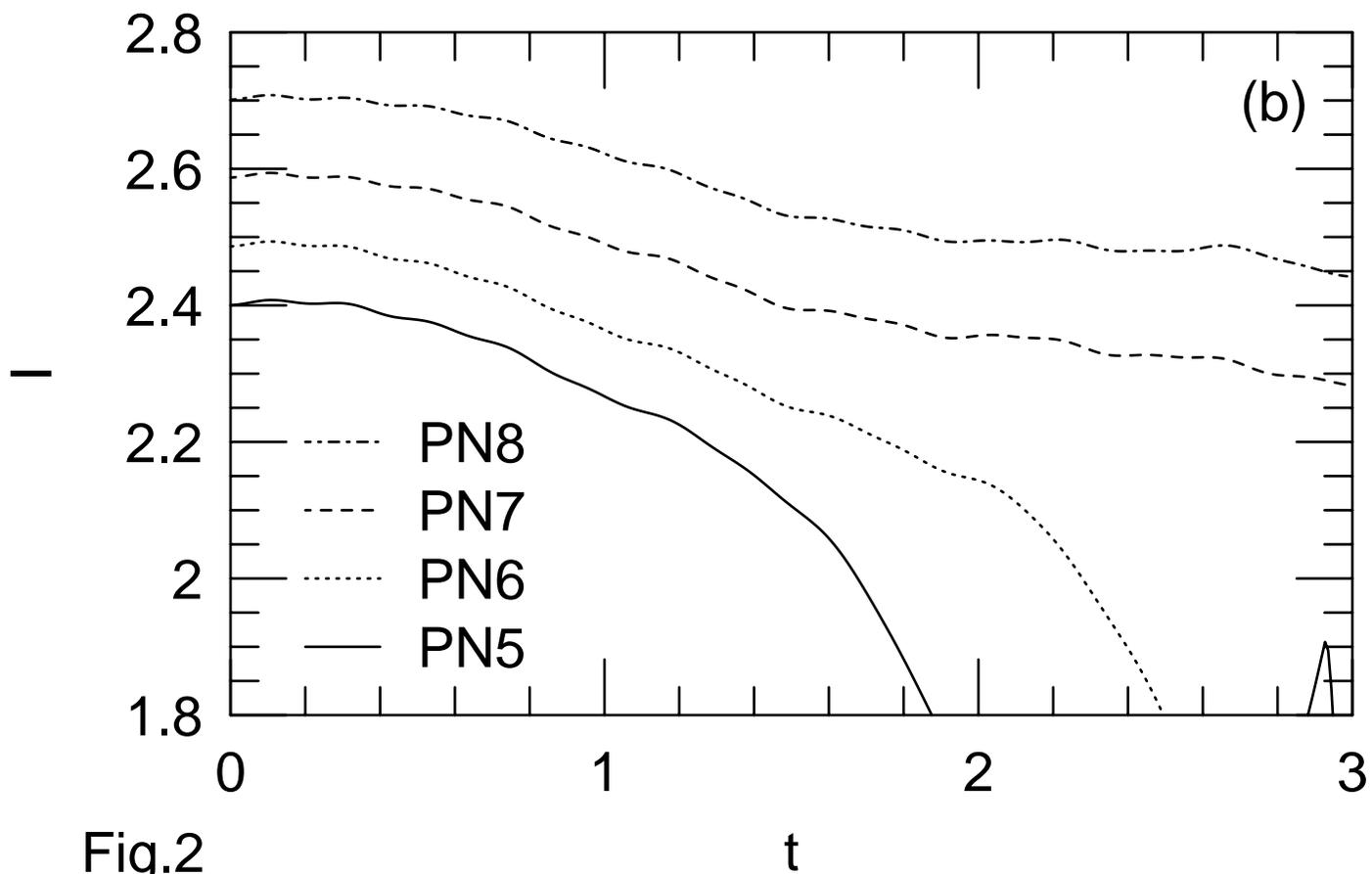
\item{\rm Fig.2} Normalized moment 
$I_{RR}/(M_*a_*^2)$ as a function of time for the Newtonian case(a) and 
the 1PN case of $C_{\rm s}\simeq 0.033$(b). Units of time is 
$4\pi\sqrt{a_*^3/GM_*}$. 
Solid, dotted, dashed and dotted-dashed lines are for 
models N5(PN5), N6(PN6), N7(PN7), and N8(PN8), respectively.
%%%%%%%%%%
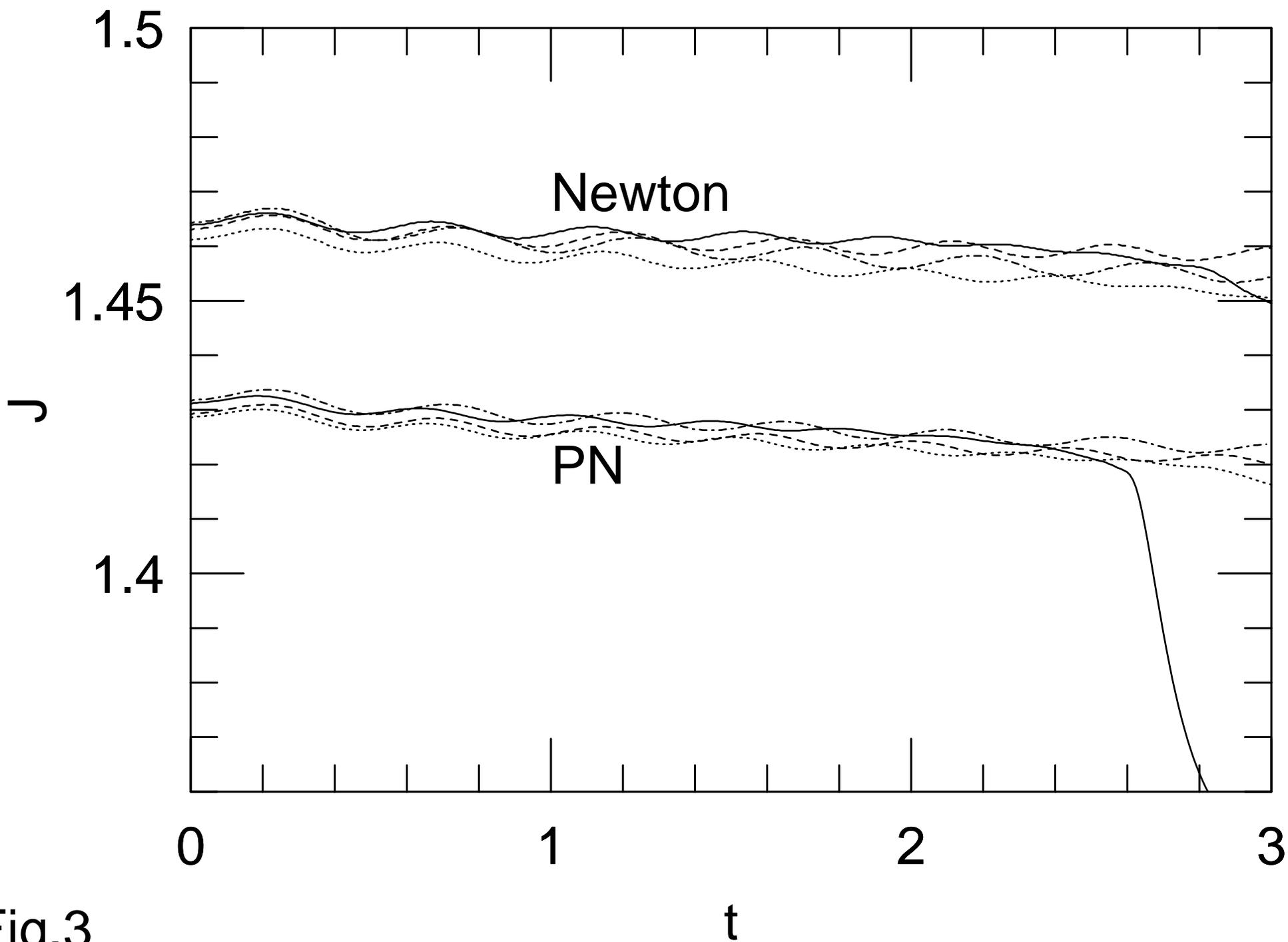
\item{\rm Fig.3} The angular momentum as a function of time. 
Units of the angular momentum and the time are 
$G^{1/2}M_*^{3/2}a_*^{1/2}$ and $4\pi\sqrt{a_*^3/GM_*}$, respectively. 
Solid, dotted, dashed and dotted-dashed lines are for 
models N5(PN5), N6(PN6), N7(PN7), and N8(PN8), respectively.
%%%%%%%%
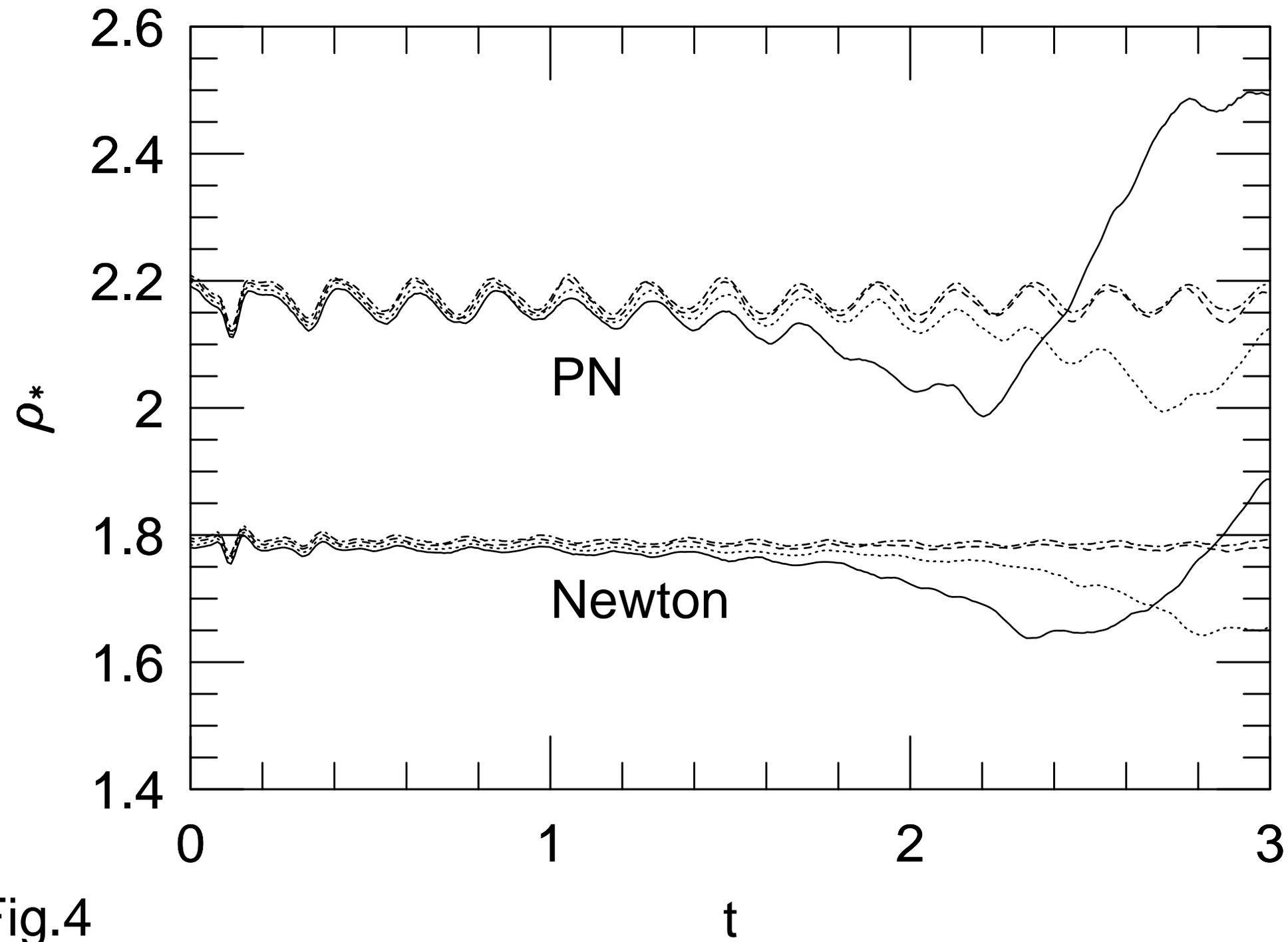
\item{\rm Fig.4} The same as fig.3, but for the maximum density. 
Units of the density is $M_*(4\pi a_*^3/3)^{-1}$. 
%%%%%%%%
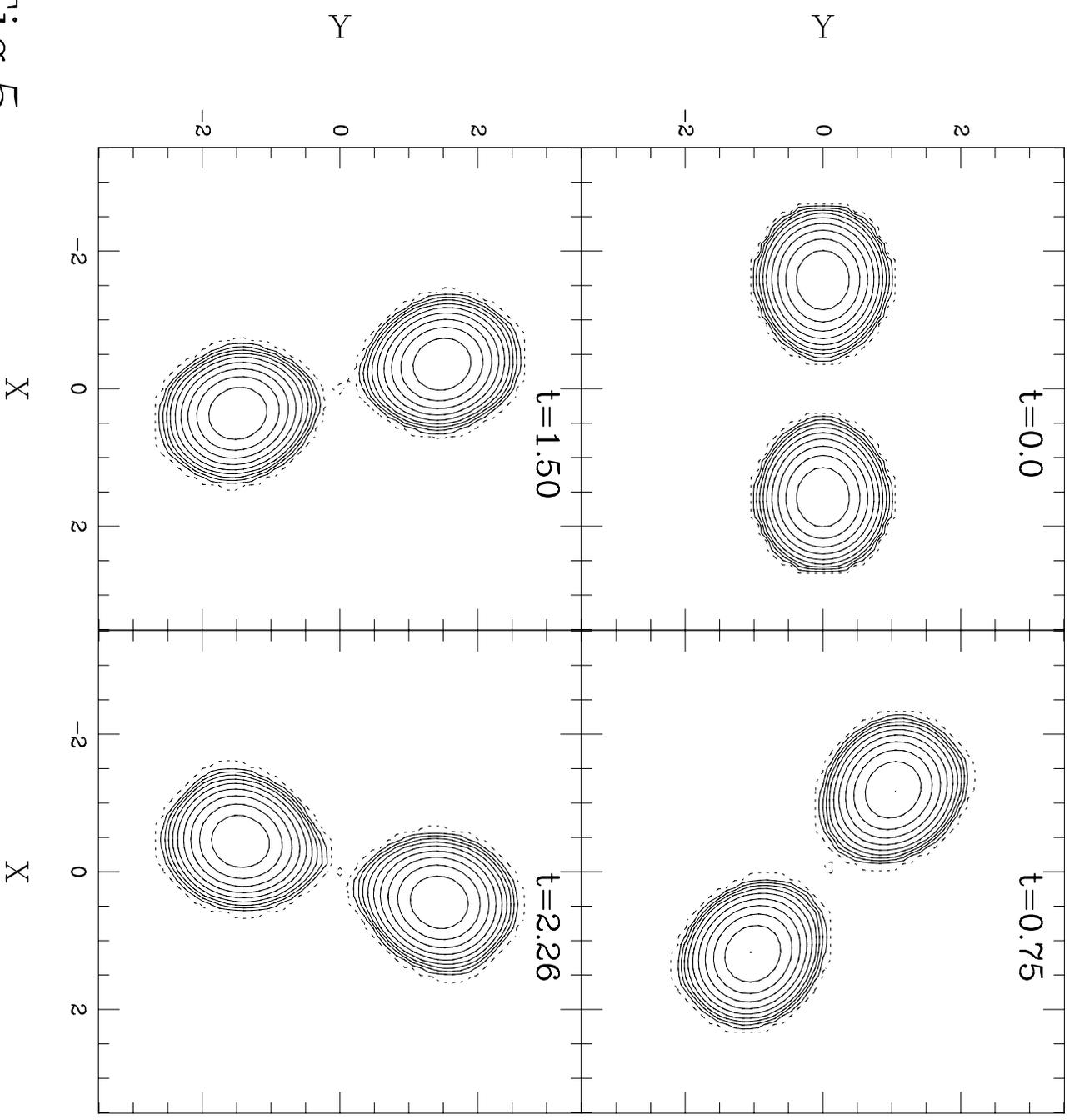
\item{\rm Fig.5} Time evolution of the density$(\rho)$ profile in the 
equatorial plane 
for model N7. $x$ and $y$-axes are shown in units of $a_*(=a_0)$. 
Note that outermost grid is placed at 4.09 in this units, and 
the orbital rotation is counterclockwise. 
Density contours(solid lines) 
are spaced with intervals of $\rho_{\rm max,0}/10$, where 
$\rho_{\rm max,0}$ is the maximum density at the initial state. 
Dotted lines denote contour of $\rho_{\rm max,0}/100$. 
%%%%%%%%
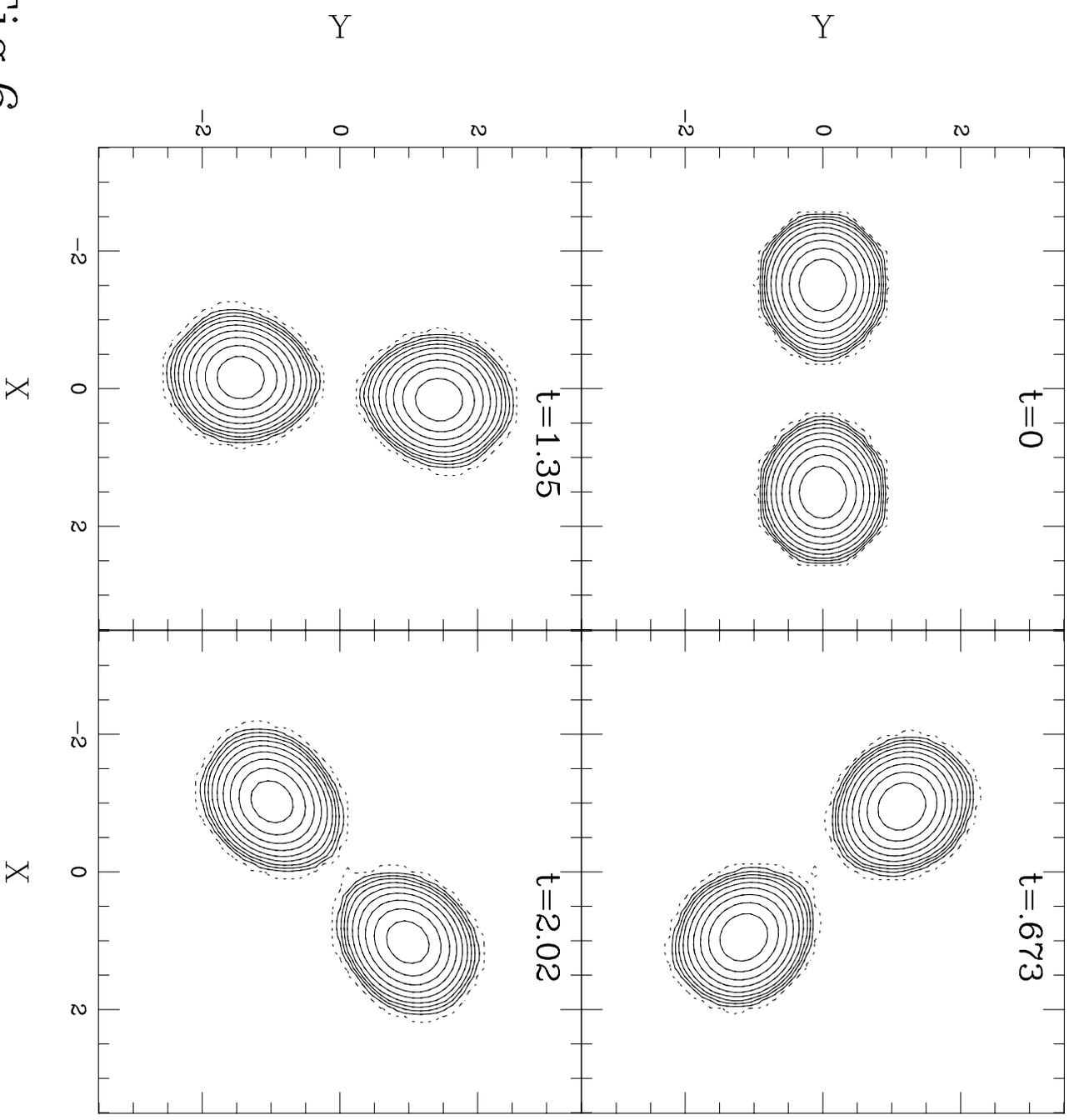
\item{\rm Fig.6} The same as fig.5, but for $\rho_*$ of model PN7. 
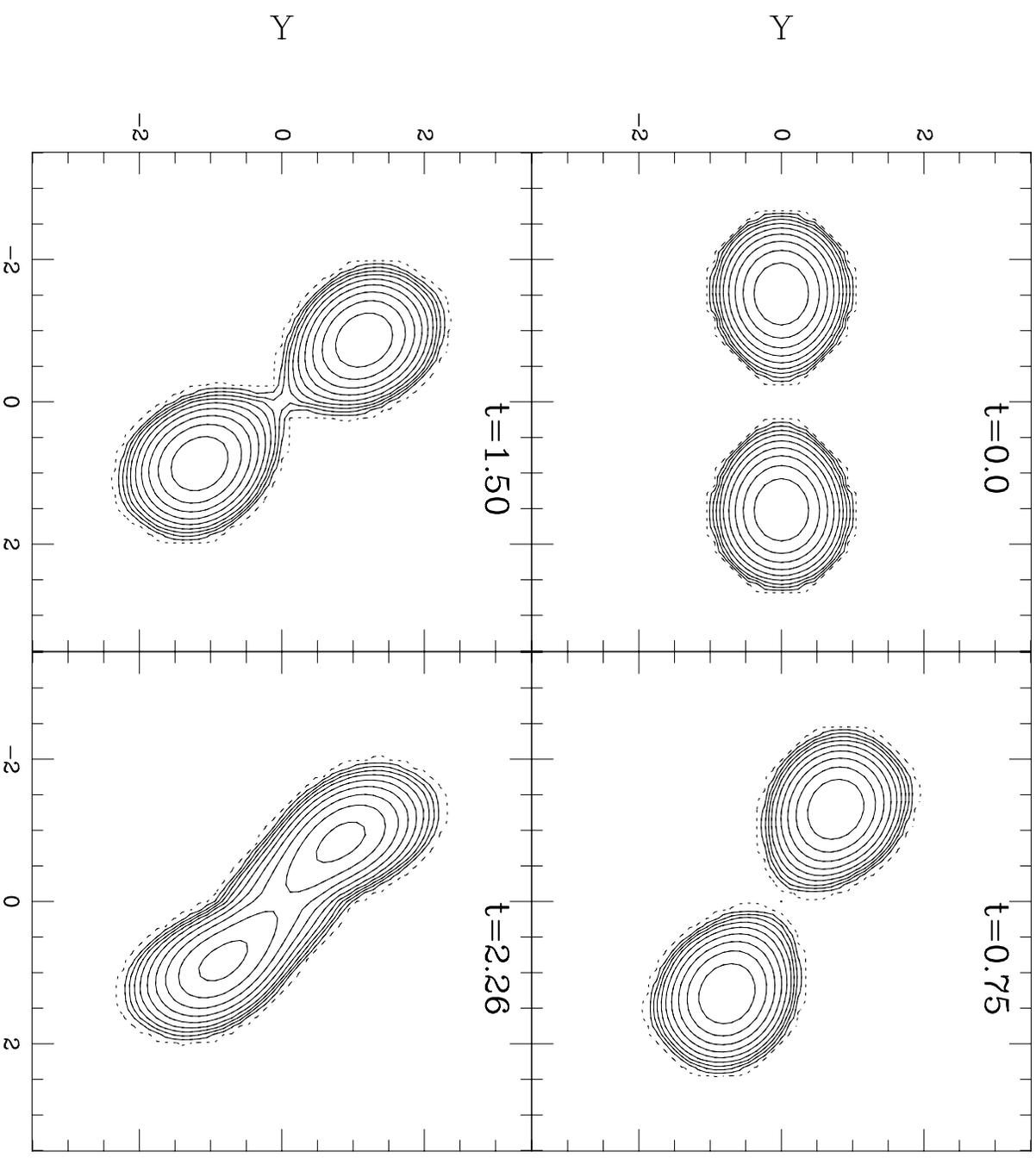
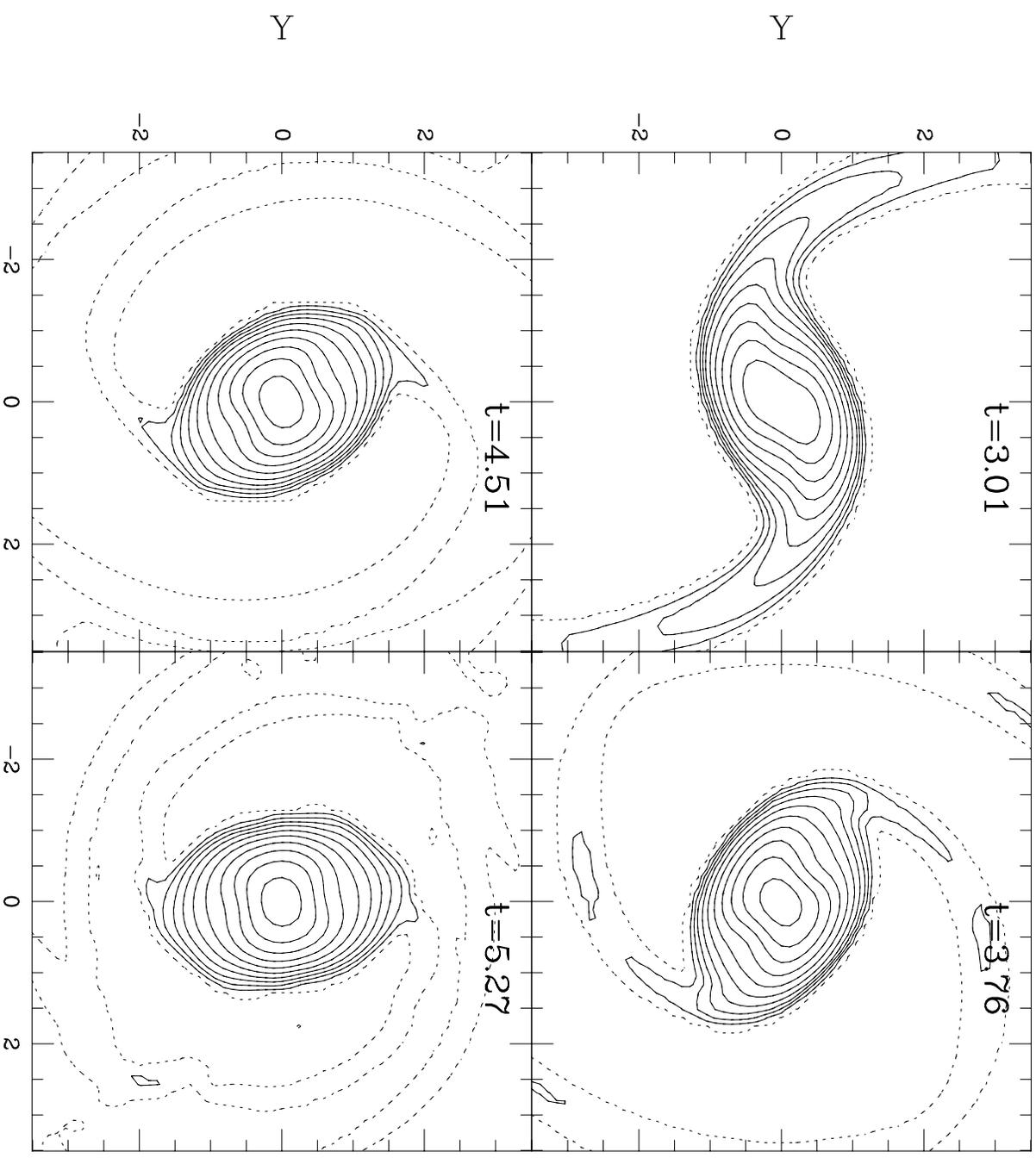
\item{\rm Fig.7} The same as fig.5, but for model N5.
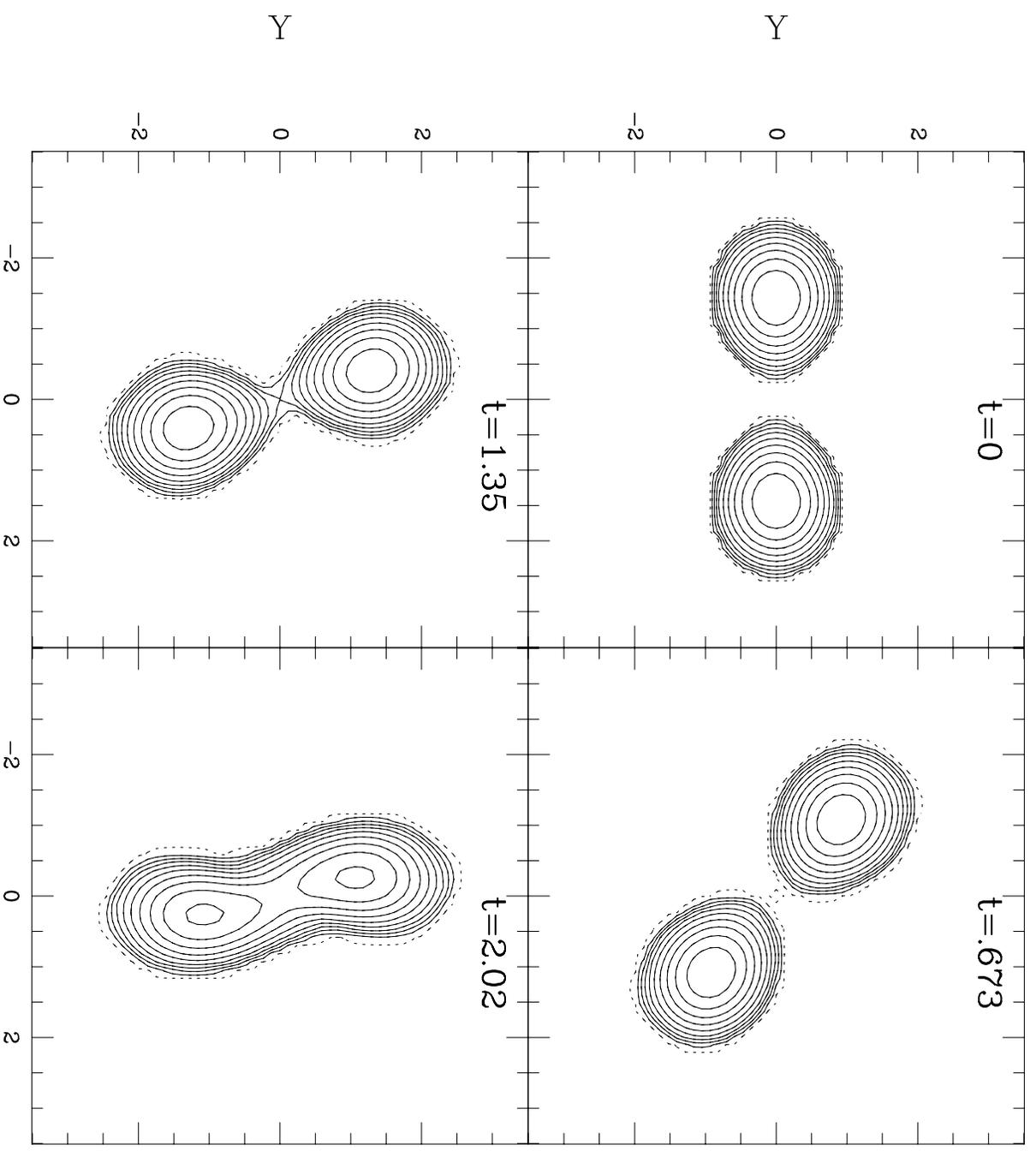
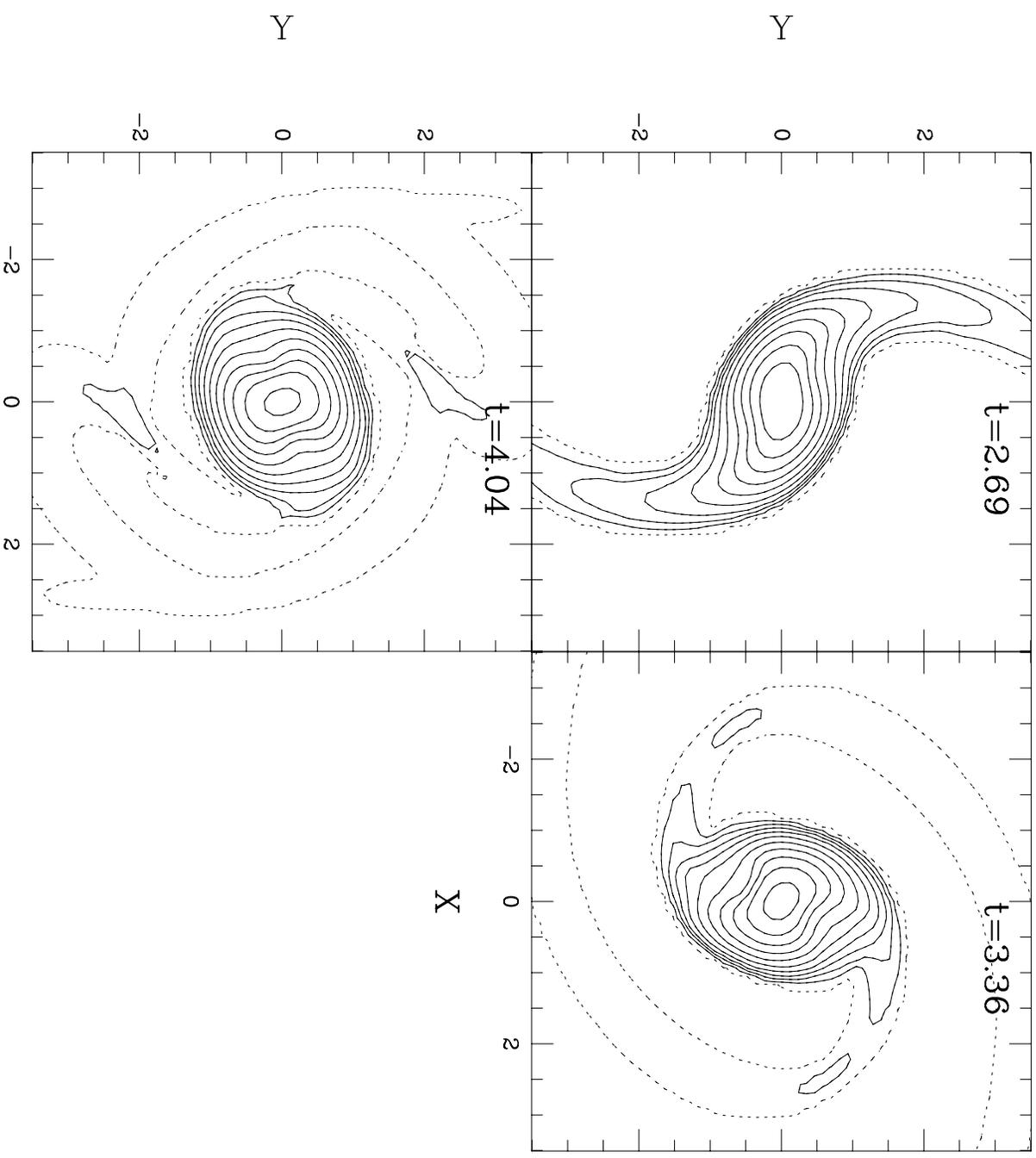
\item{\rm Fig.8} The same as fig.5, but for  $\rho_*$ of model PN5.
%%%%%%%%%%%%%
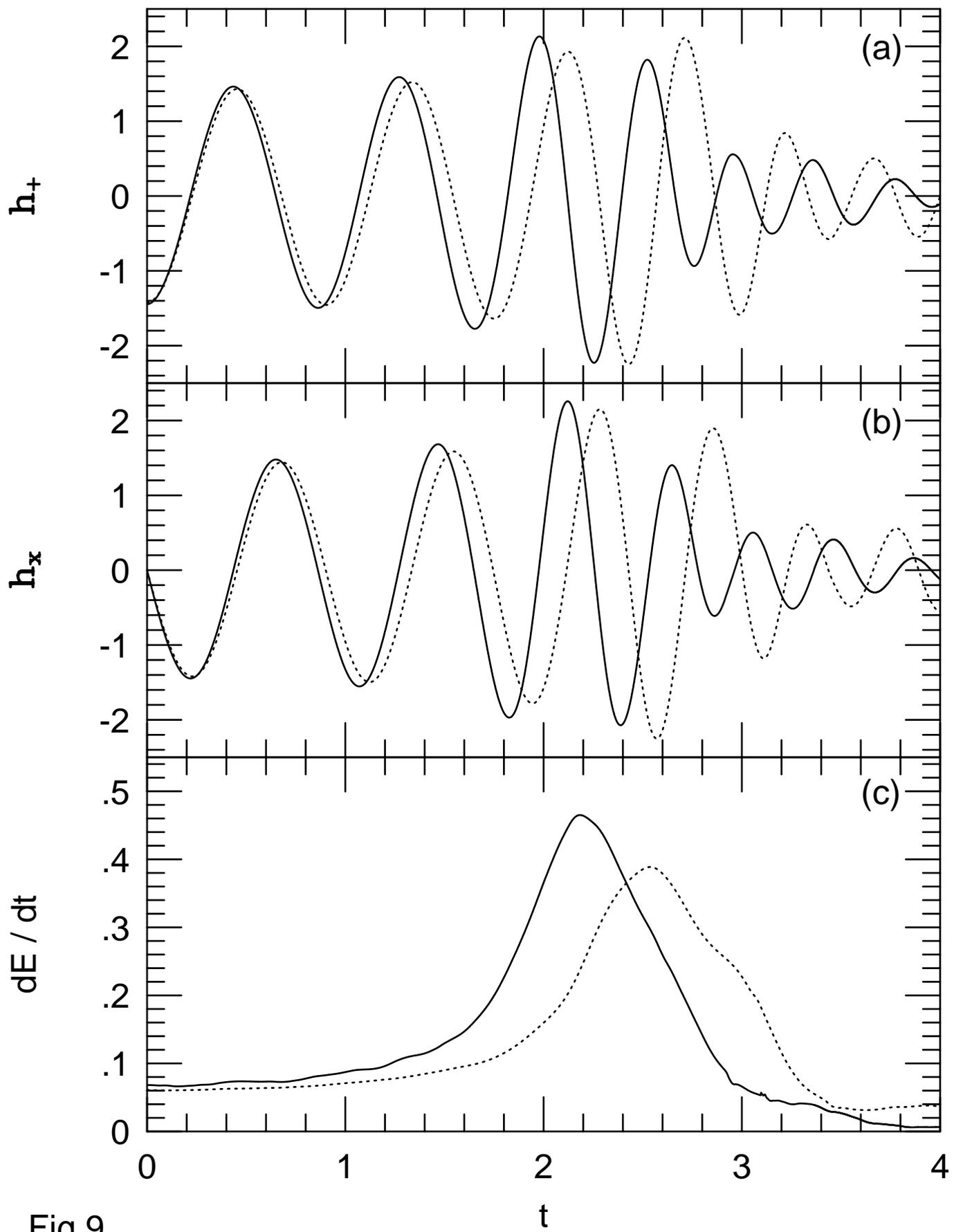
\item{\rm Fig.9} Wave forms $r_oh_+$(a) and $r_oh_{\times}$(b), 
and the luminosity(c) of gravitational waves 
for models PN5(solid lines) and N5(dotted lines). 
Units of the amplitude of $r_oh_+$ and $r_oh_{\times}$, 
the luminosity, and time are $G^2M_*^2/ c^4a_*$, 
$c^5/G(GM_*/ a_*c^2)^5$ and $4\pi\sqrt{a_*^3/GM_*}$, respectively. 
%%%
\end{description}
\end{document}